\documentclass{aa}
\usepackage{epsfig,amsfonts,amssymb}
\graphicspath{~delfosse/delfosse/article/letter_eros/}
\begin{document}
\thesaurus{10.11.1,12.04.1,08.12.2}

\title{EROS~2 proper motion survey: a field brown dwarf, 
       and an L dwarf companion to LHS~102
  \thanks{Based on observations made at the European Southern Observatory,
    La Silla, Chile.}}
\author{
B.~Goldman\inst{1},
X.~Delfosse\inst{2},
T.~Forveille\inst{3},
C.~Afonso\inst{1},
C.~Alard\inst{4},
J.N.~Albert\inst{5},
J.~Andersen\inst{6},
R.~Ansari\inst{5},
{\'E}.~Aubourg\inst{1},
P.~Bareyre\inst{1,7},
F.~Bauer\inst{1},
J.P.~Beaulieu\inst{8},
J.~Borsenberger\inst{8}\thanks{In fact, not an EROS member.},
A.~Bouquet\inst{7},
S.~Char\inst{\dag},
X.~Charlot\inst{1},
F.~Couchot\inst{5},
C.~Coutures\inst{1},
F.~Derue\inst{5},
R.~Ferlet\inst{8},
P.~Fouqu{\'e}\inst{9},
J.F.~Glicenstein\inst{1},
A.~Gould\inst{1,10},
D.~Graff\inst{10},
M.~Gros\inst{1},
J.~Haissinski\inst{5},
J.C.~Hamilton\inst{7},
D.~Hardin\inst{1,11},
J.~de Kat\inst{1},
A.~Kim\inst{7},
T.~Lasserre\inst{1},
{\'E}.~Lesquoy\inst{1},
C.~Loup\inst{8},
C.~Magneville \inst{1},
B.~Mansoux\inst{5},
J.B.~Marquette\inst{8},
E.L.~Mart{\'\i}n\inst{12},
{\'E}.~Maurice\inst{13},
A.~Milsztajn \inst{1}, 
M.~Moniez\inst{5},
N.~Palanque-Delabrouille\inst{1},
O.~Perdereau\inst{5},
L.~Pr{\'e}vot\inst{13},
N.~Regnault\inst{5},
J.~Rich\inst{1},
M.~Spiro\inst{1},
A.~Vidal-Madjar\inst{8},
L.~Vigroux\inst{1},
S.~Zylberajch\inst{1}
\\   \indent   \indent
The EROS collaboration
}
% 1 Saclay
% 2 Canaries
% 3 Grenoble 
% 4 DASGAL
% 5 LAL
% 6 Copenhaguen
% 7 College
% 8 IAP
% 9 Meudon
%10 Ohio
%11 LPNHE
%12 Berkeley
%13 Marseille

\institute{
CEA, DSM, DAPNIA,
Centre d'{\'E}tudes de Saclay, F-91191 Gif-sur-Yvette Cedex, France
\and
Instituto de Astrof{\'\i}sica de Canarias,
E-38200 La Laguna, Tenerife, Canary Islands, Spain
\and
Observatoire de Grenoble,
414 rue de la Piscine,
Domaine Universitaire de S$^{\rm t}$ Martin d'H{\`e}res,
F-38041 Grenoble,
France
\and
DASGAL, 77 avenue de l'Observatoire, F-75014 Paris, France
\and
Laboratoire de l'Acc{\'e}l{\'e}rateur Lin{\'e}aire,
IN2P3 CNRS, Universit{\'e} Paris-Sud, F-91405 Orsay Cedex, France
\and
Astronomical Observatory, Copenhagen University, Juliane Maries Vej 30,
2100 Copenhagen, Denmark
\and
Coll{\`e}ge de France, Physique Corpusculaire et Cosmologie, IN2P3 CNRS,
11 pl. M. Berthelot, F-75231 Paris Cedex, France
\and
Institut d'Astrophysique de Paris, INSU CNRS,
98~bis Boulevard Arago, F-75014 Paris, France
\and
Observatoire de Meudon,
F-92195 Meudon Cedex, France
\and
Departments of Astronomy and Physics, Ohio State University, 
Columbus, OH 43210, U.S.A
\and
LPNHE, IN2P3-CNRS-Universit{\'e}s Paris VI et VII, 4 place Jussieu, 
F-75252 Paris Cedex 05
\and
Astronomy department, University of California, Berkeley, CA 94720, U.S.A.
\and
Observatoire de Marseille,
2 pl. Le Verrier, F-13248 Marseille Cedex 04, France
}

\offprints{Bertrand.Goldman@cea.fr}

\date{Received;accepted}

\authorrunning{B. Goldman et~al.}
\titlerunning{EROS~2 proper motion survey: a brown dwarf 
              and an L dwarf companion to LHS 102}
\def\kms{{\rm km}\,{\rm s}^{-1}}
\def\kpc{{\rm kpc}}
\def\lsim{{\lesssim}}
\def\au{{\rm AU}}
\def\etal{{et~al.}}
\def\eros{{\sc eros}}
\def\macho{{\sc macho}}
\def\ie{{\em i.e.}}
\def\solar {\ifmmode_{\mathord\odot} \else $_{\mathord\odot}$\fi}% _solar
\def\Msol {\ifmmode {\,{\rm M}\solar} \else $\,{\rm M}$\solar\fi} % solar mass

\maketitle

\begin{abstract}
  We report the discovery of two L dwarfs (the new spectral class
  defined for dwarfs cooler than the M type) in a two-epoch CCD proper
  motion survey of 413~square degrees, complemented by infrared
  photometry from DENIS. One of them has a strong lithium line,
  which for very cool dwarfs is a proof of brown dwarf status.
  The other is a common proper motion
  companion to the mid-M dwarf LHS~102 (GJ~1001), which has a well
  determined trigonometric parallax.  LHS~102B is thus one of the
  coolest L dwarfs of known distance and luminosity.
\end{abstract}

\section{Introduction}
Two years ago, several very cold dwarfs were identified by DENIS
(Delfosse et~al. 1997) and Kelu~1 was found through its high proper
motion (Ruiz et~al. 1997). Follow-up observations immediately showed
that their optical spectra bear little resemblance to those of the
slightly hotter M dwarfs and resemble the previously atypical spectrum
of GD~165B (Becklin \& Zuckerman 1988; Kirkpatrick et al. 1993).
Mart{\'\i}n et~al. (1997) suggested a new class for these objects, the
L spectral class. Kirkpatrick et~al. (1999) and Mart{\'\i}n et
al. (1999) take first steps towards a definition of this class. The
main characteristic of L dwarf visible spectra, compared with those of
M dwarfs, is the gradual disappearance of the VO and TiO molecular
bands, now understood as due to depletion of titanium and vanadium
into dust. This class contains both very low mass stars with masses
just above the hydrogen burning limit and brown dwarfs, like
DENIS-P~J1228.2-1547 (Delfosse et~al. 1997) and Kelu~1 (Ruiz
et~al. 1997).

High proper motions have historically been the first tool used to
systematically search the solar neighbourhood for very low mass stars
(Luyten 1925), and the discovery of Kelu~1 shows that this remains a
powerful technique.
In this letter we report the detection of two new L dwarfs in a proper
motion survey using the EROS~2 instrument, and their confirmation by
infrared photometry from DENIS. The EROS~2 proper motion survey
primarily aims at halo white dwarfs, but a preliminary two-epochs
analysis already has useful sensitivity to very cool disk objects.
One of the new detections is a confirmed brown dwarf and the other is
a borderline object, which may be either a star or a brown dwarf. The
latter is a common proper motion companion of the parallax star
LHS~102 (d=9.6~pc, M3.5V), and significantly improves the determination
of the colour-luminosity relation for low luminosities L dwarfs.  
We first detail the observational setup and the selection process, and
then discuss the two objects in some detail.

\section{The EROS~2 proper motion survey}

\subsection{Instrument and data characteristics}
The EROS~2 two colour CCD wide-field imager (\cite{Bau97}) is mounted
at the Cassegrain focus of the 1-m Marly telescope at La Silla
(Chile).  The pixel size is 0''.6 and the field of view is
$1\,^{\circ2}$.  It contains two $8k\times{}4k$ CCD mosaics,
illuminated through a dichroic beam splitter which defines the
bandpasses.  The {\em visible} and {\em red} bands are respectively
centered close to the Johnson {\em V} and Cousins {\em I} filters, but
considerably broader.  Calibration is based on V-I=[0--1]~mag stars
and magnitudes of redder stars are indicative only.  

Proper motion
observations are performed one to two hours per dark night, within 90
minutes of the meridian to minimize atmospheric refraction.  
For the present analysis we used
$183\,^{\circ 2}$ observed close to the South Galactic Pole, and
$230\,^{\circ2}$ in the Northern Hemisphere,
which had been been observed twice or more, 
down to $V\,\simeq{}\,21.5$ and $I\,\simeq{}\,20.5$.
The experiment is expected to last until 2002, and 4 or 5
epochs will eventually be available.

\subsection{Proper motion determination}
\begin{figure} [h]
 \begin{center} \epsfig{file=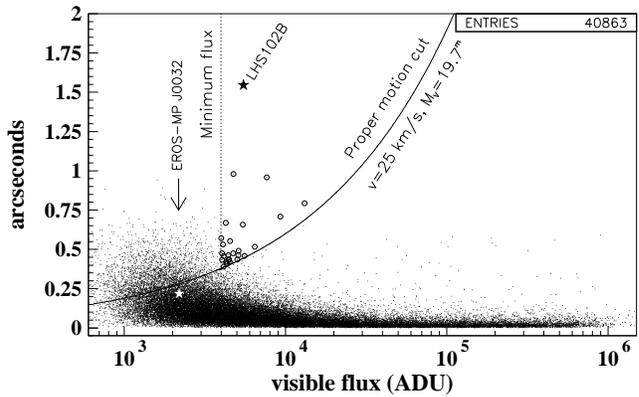,width=9cm} 
  \caption{Matching distance versus EROS flux in the visible band.
    For clarity only 5\% of the sample is displayed. The dotted
    vertical line shows the minimum accepted flux, and the solid
    curve the minimum proper motion cut.
    % and the mean position of a 25~km/s, M$_V$=19.7 star.
    The larger dots show the stars selected in the visible band only;
    only LHS~102B is selected in both bands.
}
  \label{dist}
 \end{center} 
\end{figure}

The reduction software for source detection, classification and
catalogue matching was written in the framework of the EROS PEIDA++
package (\cite{Ans96}), and processes data in (11~arcmin)$^2$ chunks.
As photon noise dominates the astrometric errors for most of the
available volume, we use a simple two-dimensional gaussian PSF fitting
to determine stellar positions.  A rough star/galaxy classification is
performed to limit galaxy contamination, with such cuts that few stars
are misclassified.  The catalogues for the two epochs are
geometrically aligned using a linear transformation adjusted to the
40~brightest stars, and matched within a search radius of
9~arcsec. The average distance between matched stars (Fig.\ref{dist})
provides an upper limit to the total astrometric error, which for
bright objects is 25~mas ($1\sigma$). For a 25~km/s disk star and a
1-year baseline this corresponds to a 5$\sigma$ detection at 40pc.

\subsection{Candidate selection}
Since our main goal is to identify dark matter contributors, we select
all objects satisfying a magnitude-dependent proper motion cut, set to
retain stars faster than V$_t$=25~km/s and fainter than $M_V=19.7^{m}$
and $M_I=16.9^{m}$, down to our detection limit (Fig.\ref{dist}).
This selection in a proper motion-magnitude diagram is mostly
sensitive to two object classes: halo white dwarfs and disk very low
mass stars and brown dwarfs. As the current analysis is based on two
epochs, we presently have to require a selection in both photometric
channels to avoid excessive contamination by spurious faint
candidates.  Even though this procedure reduces the searched volume
very significantly for L dwarfs, the one candidate it does select,
LHS~102B, has a very red EROS colour, confirmed by DENIS photometry.
Following this early success, we looked for all red objects with EROS
colour similar to LHS~102B's, regardless of their proper motion,
obtaining 25 additional candidates with $(V-I)_{EROS}\,>\;2.9$.

\begin{table}[b]
\caption{Basic parameters of the objects.  
LHS102A is satured in the EROS images, 
so its $\mu$ and $\theta_\mu$ are from \cite{Yale}. 
$\alpha$ and $\delta$ are for epoch J2000.0.
\label{tab:cand}}
\begin{center}
\footnotesize
\begin{tabular}{|c|c|c|c|}
\hline
{} & \raisebox{0pt}[13pt][7pt]{LHS102A}
 & \raisebox{0pt}[13pt][7pt]{{\bf LHS102B}}
 & \raisebox{0pt}[13pt][7pt]{{\bf EROS-MP J0032}}
 \\
\hline
{\raisebox{0pt}[12pt][6pt]{I}}
 & {\raisebox{0pt}[12pt][6pt]{$10.2 \pm 0.05$}}
 & {\raisebox{0pt}[12pt][6pt]{$17.0 \pm{} 0.05$}}
 & {\raisebox{0pt}[12pt][6pt]{$18.6 \pm{} 0.2$}} \\
\hline
{\raisebox{0pt}[12pt][6pt]{I-J}}
 & {\raisebox{0pt}[12pt][6pt]{$1.4 \pm 0.07$}} %J=8.8
 & {\raisebox{0pt}[12pt][6pt]{$3.70 \pm{} 0.07$}}
 & {$3.7 \pm{} 0.2$} \\
\hline
{\raisebox{0pt}[12pt][6pt]{J-K$_s$}}
 & {\raisebox{0pt}[12pt][6pt]{$1.0 \pm 0.07$}} %??? K=7.7
 & {\raisebox{0pt}[12pt][6pt]{$1.90 \pm{} 0.07$}}
 & {\raisebox{0pt}[12pt][6pt]{$1.25 \pm{} 0.25$}} \\
\hline
{\raisebox{0pt}[12pt][6pt]{$\mu$ ($''.\rm yr^{-1}$)}}
 & {\raisebox{0pt}[12pt][6pt]{1.618}}
 & {\raisebox{0pt}[12pt][6pt]{$1.55\pm{}0.06$}}
 & {\raisebox{0pt}[12pt][6pt]{$0.17\pm{}0.04$}} \\
\hline
{\raisebox{0pt}[12pt][6pt]{$\theta_\mu$ ($^\circ$)}} % ou PA ??
 & {\raisebox{0pt}[12pt][6pt]{154}}
 & {\raisebox{0pt}[12pt][6pt]{$158\pm{}7$}}
 & {\raisebox{0pt}[12pt][6pt]{$137\pm{}2$}} \\
\hline
{\raisebox{0pt}[12pt][6pt]{$\alpha$ (J2000)}}
 & {\raisebox{0pt}[12pt][6pt]{00:04:36.5}} %DSS 00:04:37.84 1977 J2000.
 & {\raisebox{0pt}[12pt][6pt]{00:04:33.9}} % offset de -1.6 sec tps
 & {\raisebox{0pt}[12pt][6pt]{00:32:55}} \\
\hline
{\raisebox{0pt}[12pt][6pt]{$\delta$ (J2000)}}
 & {\raisebox{0pt}[12pt][6pt]{-40:44:02}} %DSS -40:41:06.1 1977 J2000.
 & {\raisebox{0pt}[12pt][6pt]{-40:44:06}} % offset de +4 arcsec
 & {\raisebox{0pt}[12pt][6pt]{-44:05:05}} \\
\hline
\end{tabular}
\end{center}
\end{table}

\section{DENIS photometry}

\begin{figure}
\tabcolsep 0.05cm
\begin{tabular}{cc}
\psfig{height=4.3cm,file=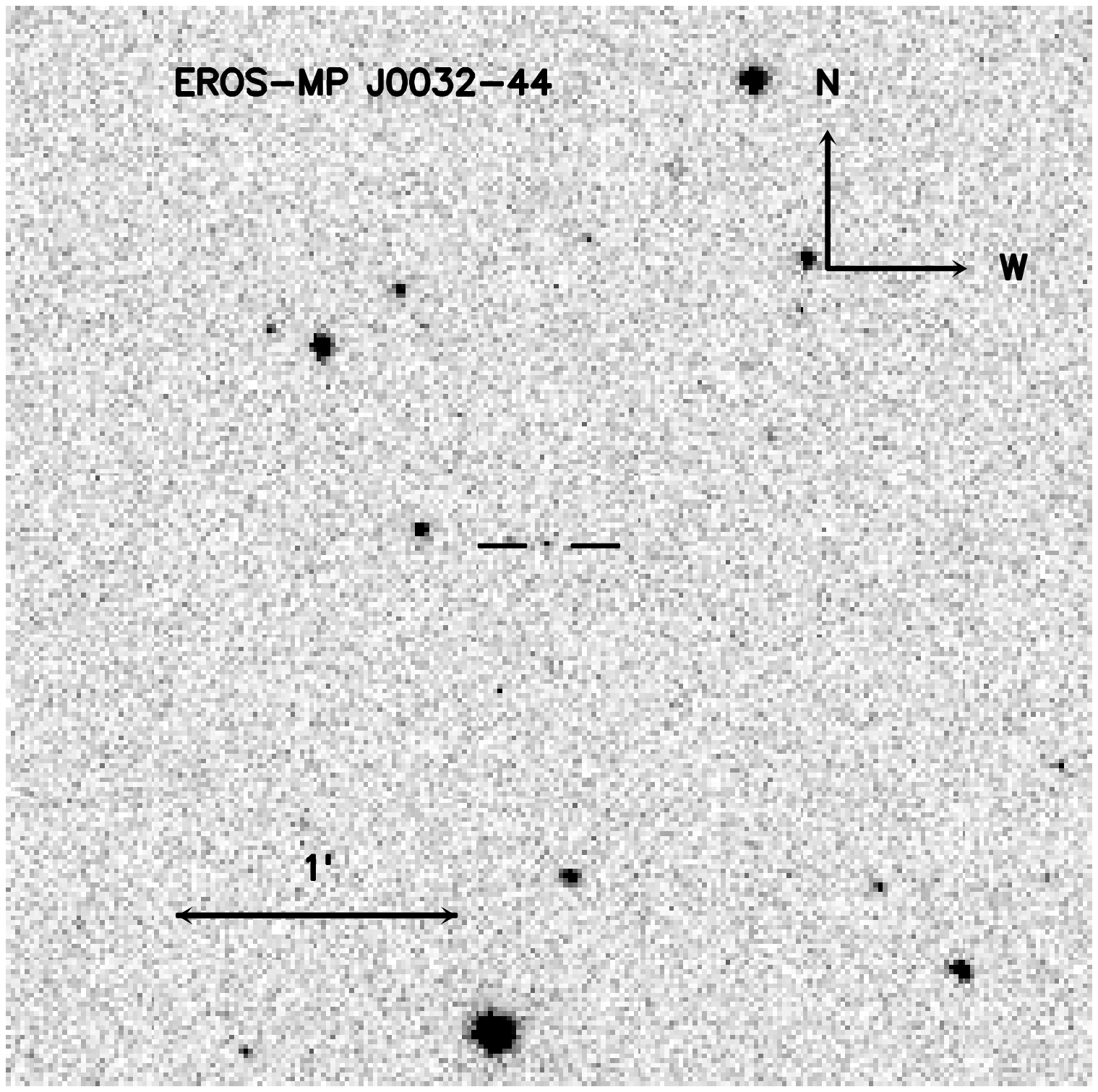,angle=-90} &
\psfig{height=4.3cm,file=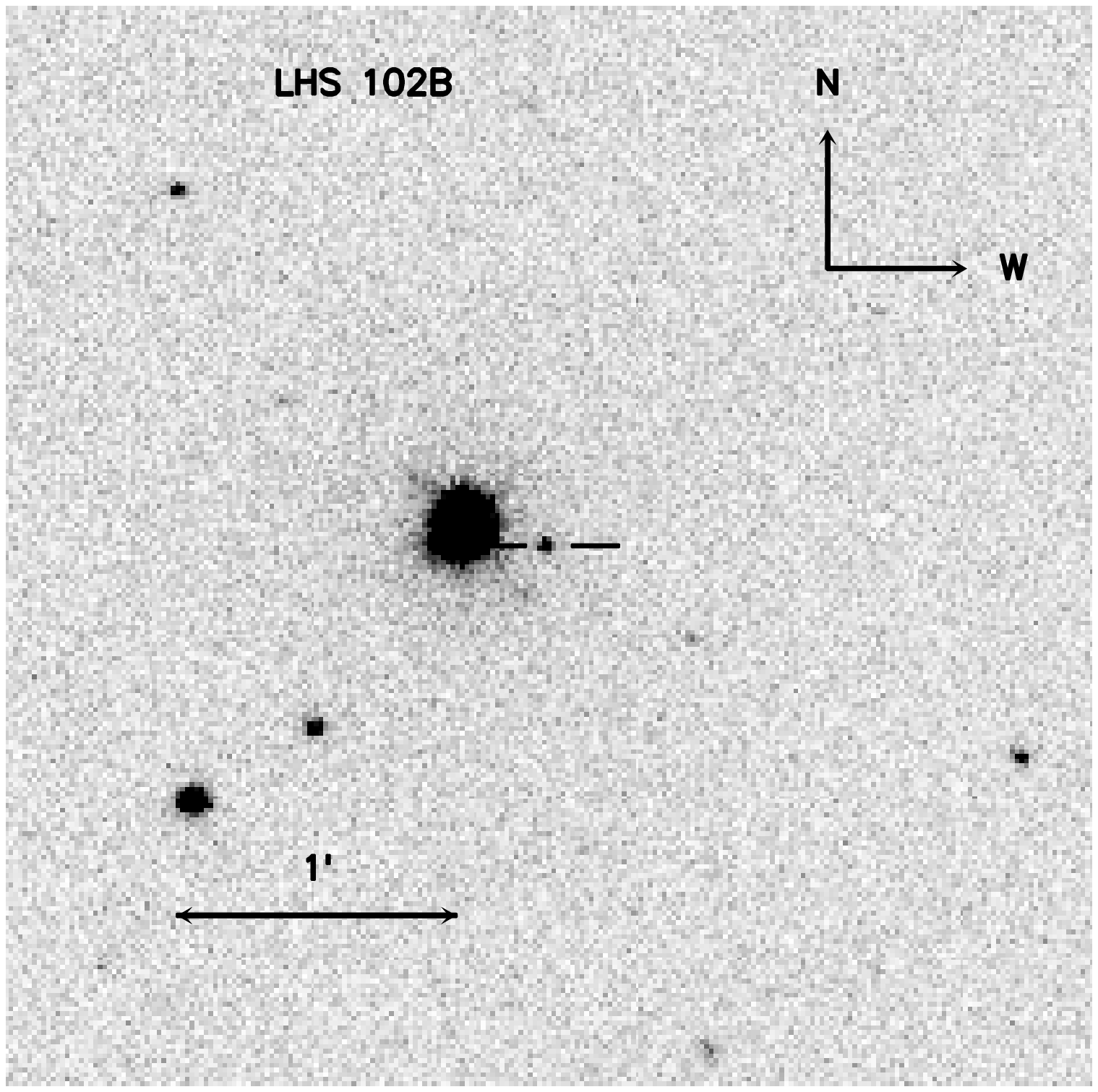,angle=-90} 
\end{tabular}
\caption{I band finding chart from the DENIS images. The chart size 
is ${\sim}~3.5'~\times~3.5$, with North up and East left.}
\label{chart}
\end{figure}

We obtained I, J and K$_s$ photometry for 12 of the 25 candidates
in the course of the DENIS survey.  DENIS observations are carried out
on the ESO 1m telescope at La Silla,
%XD: Ca fais 2 ans que je site un copet et~al. qui ne sort pas.
%peut etre mettre une autre reference maintenant.
%TF: J'ai vu quelque part (je ne sais plus ou...) qu'il etait
% accepte. Et a  dire vrai j'aimerais bien en avoir une copie, vu
% que j'en suis co-auteur et que je n'ai rien vu depuis une premiere
% version a laquelle j'avais beaucoup a redire...
%
with a three-channel infrared camera (Copet et~al. 1999). Dichroic
beam splitters separate the three channels, and focal reducing optics
provides image scales of $3''$ per pixel on the 256$\times$256 NICMOS3
arrays used for the two infrared channels and $1''$ on the
1024$\times$1024 Tektronix CCD detector of the I channel.  DENIS
photometry is an efficient estimator of spectral type or approximate
effective temperature for M and L dwarfs (Delfosse et~al. 1999). Two
of the 12 objects clearly have colours typical of L dwarfs, with I--J
larger than 3.5 (see Table \ref{tab:cand}). Fig.\ref{chart} gives
finding charts for these new L dwarfs.  High and medium resolution
optical spectra obtained at the Keck telescope confirm their L dwarf
status, and are discussed in two separate papers (\cite{Bas99},
Mart{\'\i}n et~al. 1999).

\section{Discussion}
\subsection{LHS~102B: an L dwarf companion to an M dwarf}
One of the two L dwarfs is within 20~arcsec of a previously known high
proper motion star, LHS~102 (M3.5V). It shares its proper motion of
$1.\hskip-2pt''62 \rm yr^{-1}$ towards PA=$155^\circ$, and the two
objects are thus physical companions.  The trigonometric parallax of
LHS~102 (van Altena et~al. 1995) provides a distance for the system of
$9.6\pm1.2$~pc, and LHS~102B is thus a rare case of an L dwarf of
known distance and luminosity.  Just a few months ago only two other L
dwarfs had known distances: GD~165B (Becklin \& Zuckerman 1988)
through its association with GD~165A, and Roque~25 
which Mart{\'\i n et~al. (1998b)
established to lie (at 94 \% C.L.) in the Pleiades,
whose distance $131\pm{}3\,$pc is known through main-sequence fitting 
(\cite{Pin98}, \cite{Sob98}) and
%distance utilisee par Eduardo : 127pc, ref inconnue
whose radius is $\sigma_{d}\,=\,4\pm2\,$pc (\cite{Nar99}). 
%rayon de demi-masse de aussi 3pc par (Rabout \& Mermilliot, 1997).
Kirkpatrick et al. (1999) have since
presented preliminary parallaxes for another three L dwarfs.  They are
also shown in Fig.~\ref{colour-mag}, though Mart{\'\i}n et al. (1999)
suggest that they might perhaps have problems,
as the stars would be very young for field objects ($\simeq{}0.1\,$Gyr)
and would have preserved lithium contrary to model expectations 
and observations in the Pleiades (see \cite{Mar98a}). 
It is difficult to
assess their reliability from the limited information in Kirkpatrick
et al.  (1999), but possible sources of trouble include a relatively
short timespan, and strong differential colour refraction from the
extreme colours difference between the L dwarfs and their reference
frames (as the USNO uses a very broad filter). Alternatively those
sources could be binaries, though it seems unlikely that all three
are.

Fig. \ref{colour-mag} shows M dwarfs of known distance and the six L
dwarfs in an M$_{\rm I}$ vs I--K HR diagram, together with two sets of
theoretical tracks, NextGen and NG--DUSTY.  Dust condenses in the
atmospheres of very cool dwarfs, with two main consequences: depletion
from the atmosphere of the refractory elements; such as Ti and V;
decreases line opacities; and dust continuum opacity changes the
atmospheric structure through a greenhouse effect. The NextGen
models %(Baraffe et~al. 1998) : remplace par :
(\cite{Hau99}) ignore dust condensation altogether, while the
NG--DUSTY models (Leggett et~al. 1998, \cite{All99b})
account for its effect on both the chemical equilibrium and
the continuous opacity. As can be seen in Fig. \ref{colour-mag}, the
NextGen models provide an excellent fit to near-IR colours and
luminosities of M dwarfs, but fail to reproduce the J--K reddening of
the late M and L dwarf sequence.  The NG--DUSTY models in contrast provide
an impressive fit of the near IR colours and luminosities of L dwarf,
especially when one considers the still preliminary nature of these
complex models. Clearly dust condensation plays a dominant role in
the atmospheric physics at these temperatures.

Comparison with the NG--DUSTY models gives an effective temperature of
$1700\pm50$~K for LHS~102B, consistent with that derived from the
optical spectrum (\cite{Bas99}).  The best fit is obtained for the
5~Gyr isochrone and a mass of $0.072\Msol$ (just at the
stellar/substellar mass limit for models using NG--DUSTY atmospheres
(\cite{All99b})), but the data are also consistent with a 1~Gyr age
and a (substellar) mass just above $0.06\Msol$. Since LHS~102B has
depleted its lithium (\cite{Bas99}), its mass must be larger than
$0.06\Msol$ and its age therefore cannot be less than $\sim$1~Gyr.
The optical spectrum indicates shows weak H$_{\alpha}$ emission, and
this low level chromospheric activity may indicate that LHS~102B is
not very much older than that minimum 
age.  It can be either a star or a brown dwarf and we cannot presently
say on which side of the border it stands.

%We have obtained a high accurate radial velocity of LHS~102A
%(V$_{\rm r}~=~17.7\pm0.1{\rm km.s}^{-1}$) with the 1.2-m
%new Swiss Telescope at La Silla (Chile), using the spectrograph
%CORALIE. This velocity combined with the proper motion of van Altena et
%al. (1995), give a UVW space motion (U=-4, V=-75, W=-8) wich is coherent
%with a disk kinematic population.

\begin{figure*}
\tabcolsep 0cm
\begin{tabular}{ccc}
\psfig{height=8.7cm,file=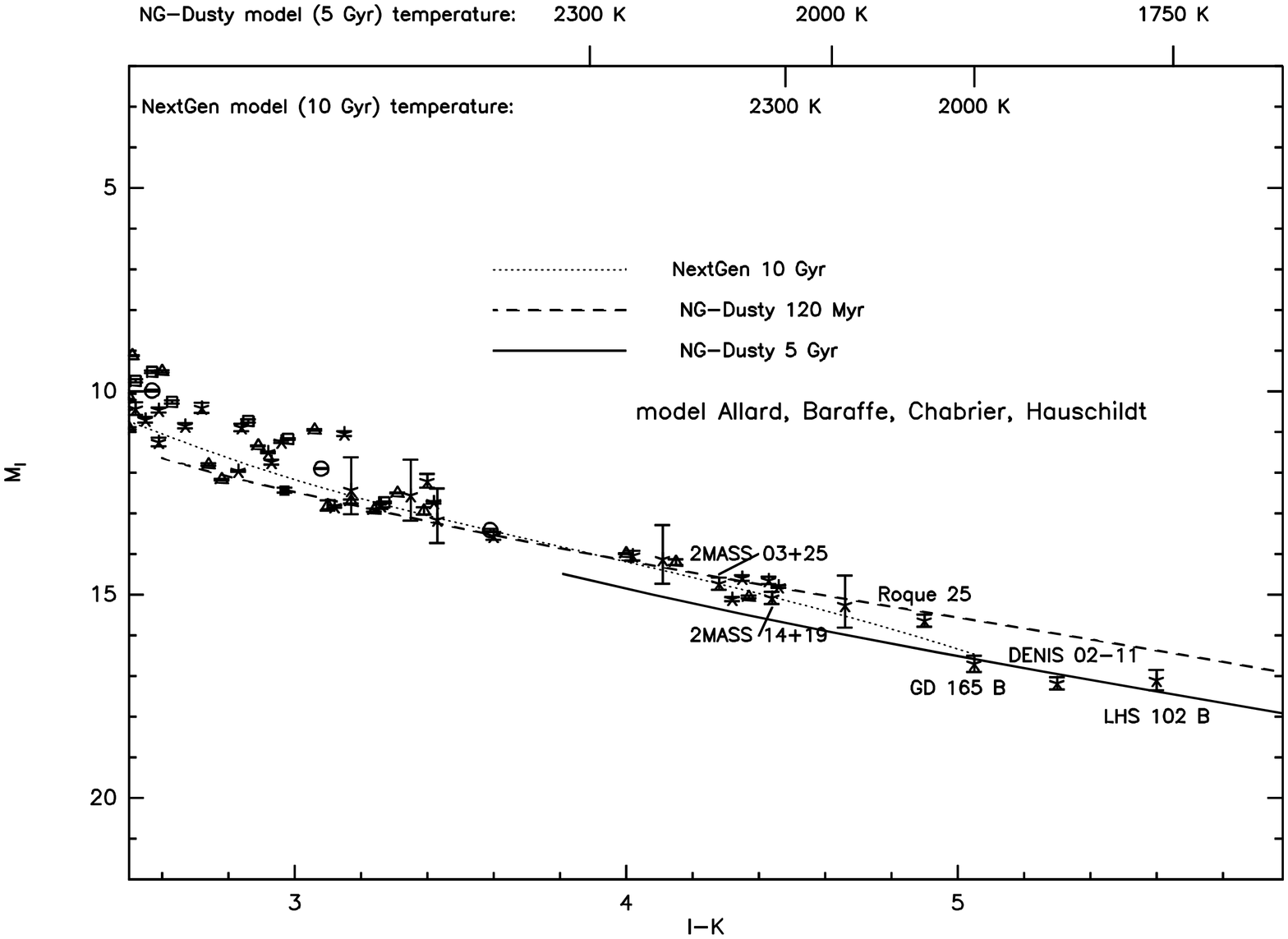,angle=-90} & \hspace{0.6cm} &
\psfig{height=8.7cm,file=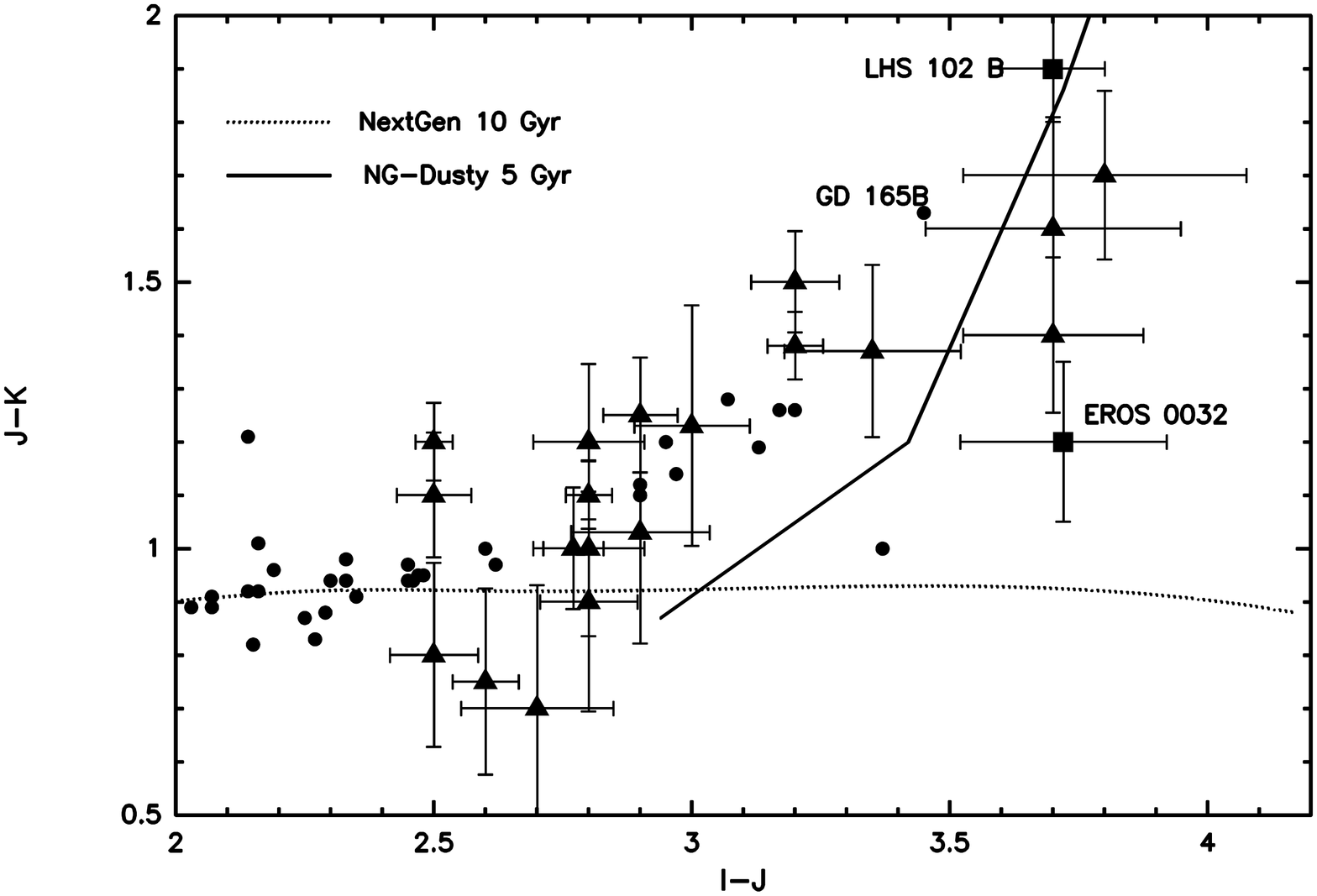,angle=-90} 
\end{tabular}
\caption{a) M$_{\rm I}$:I--K 
HR diagram for our objects, along with M and L dwarfs with known
distances (from Leggett 1992, Tinney et~al. 1993 and Kirkpatrick
et~al. 1999).
%{\em IL FAUT DECRIRE QUI EST QUOI DANS LES MODELES}
%ENS-Lyon 
Models are overlaid for both dust-free and NG--DUSTY atmospheres
and for ages of 5~Gyr ($\sim$appropriate for field objects) and
120~Myr (appropriate for the Pleiades brown dwarf Roque~25).  b)
Colour-colour diagram of M and L dwarfs (Fig. 6 of Delfosse et~al. 
1999) with the two new L dwarfs (square symbols).
%The triangle are the DENIS mini-survey objects and the
%circle the precedently known M dwarfs.
}
\label{colour-mag}
\end{figure*}

\subsection{EROS-MP J0032-4405: a field brown dwarf}
The second object, EROS-MP J0032-4405, has I--J$=3.7\pm0.2$ and
J--K$=1.2\pm0.14$. From comparison with NG--DUSTY atmospheric models,
we obtain an effective temperature of $T_{eff}=1850\pm150$~K.  
This is only marginally consistent with the effective temperature of
2200+-100~K, which corresponds to the L0 spectral type derived by \cite{Mar99}, 
with 1-$\sigma$ error bars extending to M9.5--L0.5 classes. 
The 670.8 nm lithium line absorption in the
optical spectrum (see Fig.\ref{spectre} and Mart{\'\i}n et al. 1999) 
indicates that this fully
convective very cool dwarf has not depleted its lithium. Since lithium
is destroyed by proton capture at lower temperature than needed for
hydrogen fusion (Rebolo et al. 1992), EROS-MP J0032-4405 has to be a
brown dwarf, less massive than 0.06\Msol.  Since models show that
0.06\Msol\,brown dwarfs cool down to effective temperatures of
$\sim$~1800~K at an age of $\sim$1Gyr (and less massive ones cool
faster), it must also be younger than $\sim$1~Gyr.

\begin{figure} [h]
 \begin{center} \epsfig{file=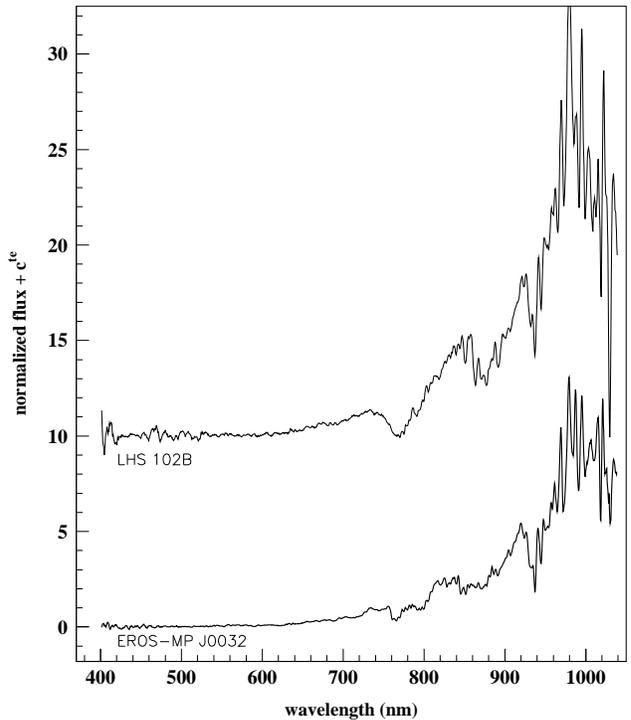,width=9cm} 
 \caption{Spectrum of EROS~J0032-4405 and LHS~102B 
          (shifted by 10~units), as in \cite{Mar99}.
          The flux has been normalized to the counts in the region
          738--742~nm. See \cite{Mar99} for details.
         }
  \label{spectre}
 \end{center} 
\end{figure}

\section{Conclusions}

Coming after the recent discoveries of DENIS-P J1228.2-1547 (Delfosse
et al. 1997), Kelu~1 (Ruiz et~al. 1997), LP~944-20 (Tinney 1998) and
of field brown dwarfs found by 2MASS (Kirkpatrick et al. 1999), the
identification of EROS-MP J0032-4405 in a small surveyed volume again
indicates that brown dwarfs are quite common in the solar
neighbourhood. This should not come as a surprise, since the mass
function of the Pleiades cluster (Bouvier et al. 1998; Zapatero Osorio
et al. 1999) rises mildly into the brown dwarf domain.
%
%With a I magnitude of I=$18.6\pm0.2$, and using the colour-luminosity of the
%figure \ref{colour-mag}, the photometric distance of EROS-MP~J003255-440505
%is $28\pm8$~pc.
%La distance est calculee en prenant Mi16.5+-0.5, suivant les modeles des
%Chabrier et al. d'ou (M-m)i=2.1+-2.1

EROS~2 combined with DENIS has proved their capabilities to find L
dwarfs in the solar neighbourhood.
We are currently analysing a larger area
with three epochs and a longer time baseline. This will allow us
to relax the requirement of a detection in the two colour channels,
and provide more accurate proper motions. We therefore hope to find
new, cooler objects, and to further characterize the brown dwarf
population of the solar neighbourhood.

\begin{acknowledgements}
We are grateful to Daniel Lacroix and the technical staff at the
Observatoire de Haute Provence and to Andr{\'e} Baranne for their help in
refurbishing the MARLY telescope and remounting it in La Silla. We
also thank the technical staff of ESO La Silla for their support of
EROS. We thank Jean-Fran{\c c}ois Lecointe for assistance with the
online computing. We are grateful to Gilles Chabrier, Isabelle
Baraffe and France Allard for useful discussions, 
and for communicating unpublished results,
and to Harmut Jahreiss for pointing out an error in LHS102 positions.
B.G. would like to thank the whole
staff of Universidad de Chile, Santiago, and ESO--Chile, for support
while at Cerro Cal{\'a}n. 
The DENIS project is supported by the {\it SCIENCE and the Human
Capital and Mobility} plans of the European Commission under grants
CT920791 and CT940627, by the French INSU, the Minist{\`e}re de
l'Education Nationale and the CNRS, in Germany by the State of
Baden-W{\"u}rtemberg, in Spain by the DGICYT, in Italy by the CNR, by
the Austrian Fonds zur F{\"o}rderung der wissenschaftlichen Forschung
und Bundesministerium f{\"u}r Wissenschaft und Forschung, in Brazil by
the Fundation for the development of Scientific Research of the State
of S{\~a}o Paulo (FAPESP), and by the Hungarian OTKA grants F-4239 and
F-013990, and the ESO C \& EE grant A-04-046.

\end{acknowledgements}

\end{document}